\documentclass{iopart}

\usepackage{color}
\usepackage{graphicx}
\usepackage{xspace}
\usepackage[utf8]{inputenc}
\newcommand{\x}{\,$\times$\,}

\begin{document}

\title[Non-linear vortex gyration]{Observation of non-linear magnetic vortex gyration by X-ray microscopy and micromagnetic simulations}

\author{A. Vansteenkiste$^1$, B. Van de Wiele$^2$, M. Weigand$^3$, H. Stoll$^3$, K.W. Chou$^4$, T. Tyliszczak$^4$, G. Woltersdorf$^5$, C. H. Back$^5$, G. Sch{\"u}tz$^3$, B. {Van Waeyenberge} $^1$}
\address{
  $^1$ Department of Solid State Sciences, Ghent University, Kortrijksesteenweg 281-S1, 9000 Gent, Belgium.\\
  $^2$ Department of Electrical Energy, Systems and Automation, Ghent University, Sint Pietersnieuwstraat 41, B-9000 Ghent, Belgium.\\
  $^3$ Max-Planck-Institut für Intelligente Systeme (ehemals Max-Planck-Institut für Metallforschung), Heisenbergstr. 3, 70569 Stuttgart \\
  $^4$ Advanced Light Source, LBNL, 94720 Berkeley, CA, USA.\\
  $^5$ Institut f{\"u}r Experimentelle und Angewandte Physik, Universit\"at Regensburg, 93040 Regensburg, Germany.\\
}
\date{\today}

\begin{abstract}
Magnetic vortex gyration in a 500\,nm wide, 50\,nm thick Permalloy square platelet was investigated using time-resolved scanning transmission x-ray microscopy and micromagnetic simulations. The response of the vortex core on an in-plane oscillating magnetic field was studied as a function of the excitation frequency and amplitude. Non-linear behavior was observed in the form of a \emph{redshift} of the gyrotropic resonance. I.e., when the excitation amplitude was increased, the resonance frequency was found to decrease. The nonlinearity is in agreement with extensive micromagnetic simulations. This work complements previous reports on thinner nanostructures where a frequency \emph{blueshift} was observed.
\end{abstract}


\maketitle

\section{Introduction}

A magnetic vortex configuration occurs in, e.g., disk- and square-shaped soft magnetic nanostructures of suited size \cite{raabe00, zhu05}. This configuration exhibits an in-plane curling magnetization which forms a flux-closure (see Fig. \ref{bending}). In the case of a square platelet structure, four domains separated by 90$^\circ$ Néel walls are formed, while no domain walls occur in disk-shaped structures. At the center of the vortex, an out-of-plane magnetized vortex core appears. The magnetization of the core has two possible directions (``up'' or ``down''), which could enable vortices to be used as magnetic memory bits \cite{vanwaeyenberge06}.\\

The vortex has a characteristic excitation mode: the vortex gyration mode \cite{argyle84}. This mode corresponds to a circular motion of the vortex core around the center of the structure. It can be excited with, e.g., an in-plane alternating \cite{chou07} or rotating \cite{curcic08} magnetic field, by magnetic field pulses \cite{weigand09} or directly by an electrical current \cite{bolte08}. The gyrotropic mode currently receives considerable attention, since sufficient excitation causes the vortex core polarization to switch its direction \cite{vanwaeyenberge06}. This happens whenever the motion of the vortex core is accelerated beyond a threshold velocity in the order of 300\,m/s \cite{yamada07, vansteenkiste09-nphys}, independent of the excitation type. Alternating magnetic fields with a frequency close to the gyrotropic frequency require only modest amplitudes of the order of one mT to cause vortex core switching \cite{vanwaeyenberge06, curcic08}. A precise determination of this frequency is thus a prerequisite to low-power vortex core memories.\\

Subjected to a small excitation, the vortex gyration behaves linearly: the dynamics can be  well described by a two-dimensional harmonic oscillator model \cite{thiele73, huber82, novosad05, kruger07}. Such theoretical models describe the vortex core as a rigid quasi particle that moves in a harmonic potential. The resulting gyration radius as a function of the excitation frequency is a typical damped harmonic oscillator curve, showing a single peak at the eigenfrequency which is independent of the excitation amplitude.\\

At large excitations, on the other hand, these harmonic models break down and the gyration dynamics show nonlinear behavior. Argyle was the first to investigate the gyrotropic resonance and already noticed such nonlinearity in 1984 \cite{argyle84}. His group investigated a 15\,$\mu$m wide garnet structure. The optical Cotton-Mouton effect was used to image the displacement of the domain walls while the vortex was excited by in-plane alternating fields. When the excitation frequency was scanned, a peak response was found at the gyrotropic eigenfrequency. When the same scan was repeated at a higher excitation amplitude however, the resonance peak had moved towards a lower frequency and also became broader. This frequency ``redshift'' was attributed to nonlinear dynamics at large excitation amplitudes, without further investigation. 
More recently, Ferromagnetic Resonance (FMR) techniques have been employed \cite{buchanan07}. In such studies, an ensemble of presumably identical nanostructures is positioned on top of an electrical strip line through which an alternating current is sent. This induces an in-plane alternating magnetic field in the radio frequency range, which excites the gyrotropic mode. The resonance is then studied by measuring the microwave absorption as a function of the excitation frequency and amplitude. However, such experiments can not determine whether vortex core switching occurs, making the interpretation difficult. \\

In this work, a more direct study of the frequency- and field dependence of the gyrotropic resonance in Permalloy nanostructures is presented. Experimental investigation with time-resolved magnetic Scanning Transmission X-ray Microscopy (STXM) allowed for a single nanostructure, rather than an array, to be investigated. The dynamics of the vortex core was imaged directly, allowing its position and velocity to be determined. As vortex core switching could be detected, the effects of switching could clearly be separated from the non-linearities in the dynamics.\\ 

Additionally, micromagnetic simulations were carried out on Graphical Processing Units using the in-house developed \textsc{MuMax} software \cite{mumax}. The speedup delivered by this hardware allowed an elaborate scan of the frequency/amplitude parameter space. Previous micromagnetic studies like \cite{lee07-apl} included only a relatively limited number of simulations.\\

Finally, previous works \cite{lee07-apl, Guslienko2010} focused on relatively thin nanostructures, in which case the vortex core can easily approach the nanostructure boundaries. This was found to lead to a frequency \emph{blueshift} at high excitation amplitudes --- in contrast to the \emph{redshift} found in the thicker nanostructures studied here.


\section{Experiment}
\subsection{Experimental method}

One element of an array of nominally 500\,nm$\times$\,500\,nm\,$\times$\,50\,nm square nanostructures, patterned in an evaporated Permalloy film was investigated with the Scanning Transmission X-ray Microscope at the Advanced Light Source (ALS, beamline 11.0.2) \cite{kilcoyne03}. By orienting the sample plane perpendicular to the photon beam and using the X-ray magnetic circular dichroism (XMCD) effect \cite{schutz87}, the out-of-plane component of the magnetization was measured. This allowed for the vortex core to be directly imaged and its polarization to be determined \cite{chou07}.\\

The nanostructures were positioned on top of a 2.5\,$\mu$m wide, 150\,nm thick Cu strip line. To excite the gyrotropic mode, an in-plane alternating magnetic field $B(t)=B_0\sin\left(2\pi f t\right)$, was induced by sending an alternating current through the strip line. By taking advantage of the short, intense X-ray flashes provided by the synchrotron, stroboscopic images of the vortex core during a steady-state gyration could be recorded. In this experiment, the synchronization requirements between the acquisition and the synchrotron ring allowed only excitation frequencies $f = n \times $\,20\,MHz with $n$ no multiple of 5. Per period of the excitation, 25 images were acquired. In order to collect sufficient photon counts, the responses of many gyration periods were summed up, and consequently only a steady-state motion could be recorded. Non-magnetic and non-dynamic contributions to the contrast were removed by subtracting the average of all frames. The resulting images show the gyrating vortex core as a white or black spot, depending on its polarization (see Fig. \ref{fig1}). \\

Since the vortex core was imaged directly, vortex core switching could be detected. When the magnetic excitation amplitude is sufficiently increased, the vortex core starts toggling back and forth between the two possible polarizations which causes the recorded contrast in the experiment to vanish entirely (as only steady-state dynamics can be recorded). In such a case, the switching could be confirmed by again lowering the excitation amplitude and determining the core polarization. In about half of the cases, the polarization was found to have switched, which can be expected when the core has been constantly toggling back and forth for a long time \cite{yamada07}. Also, each time the vortex core had switched, a short magnetic pulse was employed to switch it back to its original polarization \cite{weigand09} before the experiment was continued. In this way, all data was recorded for a vortex core pointing up, and possible effects due to the small asymmetry between the core polarizations \cite{chou07, curcic08, vansteenkiste09-njp} were excluded.\\ 

We have excluded the possibility that the observed changes in gyrotropic frequency (see further) could be caused by sample heating due to the electrical excitation. Sufficient heating could cause a frequency redshift due to a corresponding reduction of the saturation magnetization. E.g., Kamionka \textit{et al.} found about a 5\% decrease of resonance frequency when heating the sample by 100$^\circ$C \cite{Kamionka2011}. Therefore, we have determined the temperature of the copper stripline by accurately measuring its electrical resistance while sending different currents through. Even at the highest currents used in the experiment, the temperature increase remained below 1$^\circ$C, so it can not be responsible for the observed non-linearities. Furthermore, the micromagnetic simulations presented in section \ref{simulations} were carried out at $T=0$ and reproduce the non-linearities.\\

\subsection{Experimental results}

\begin{figure}[!htb]
\centering
\includegraphics[width=0.7\linewidth]{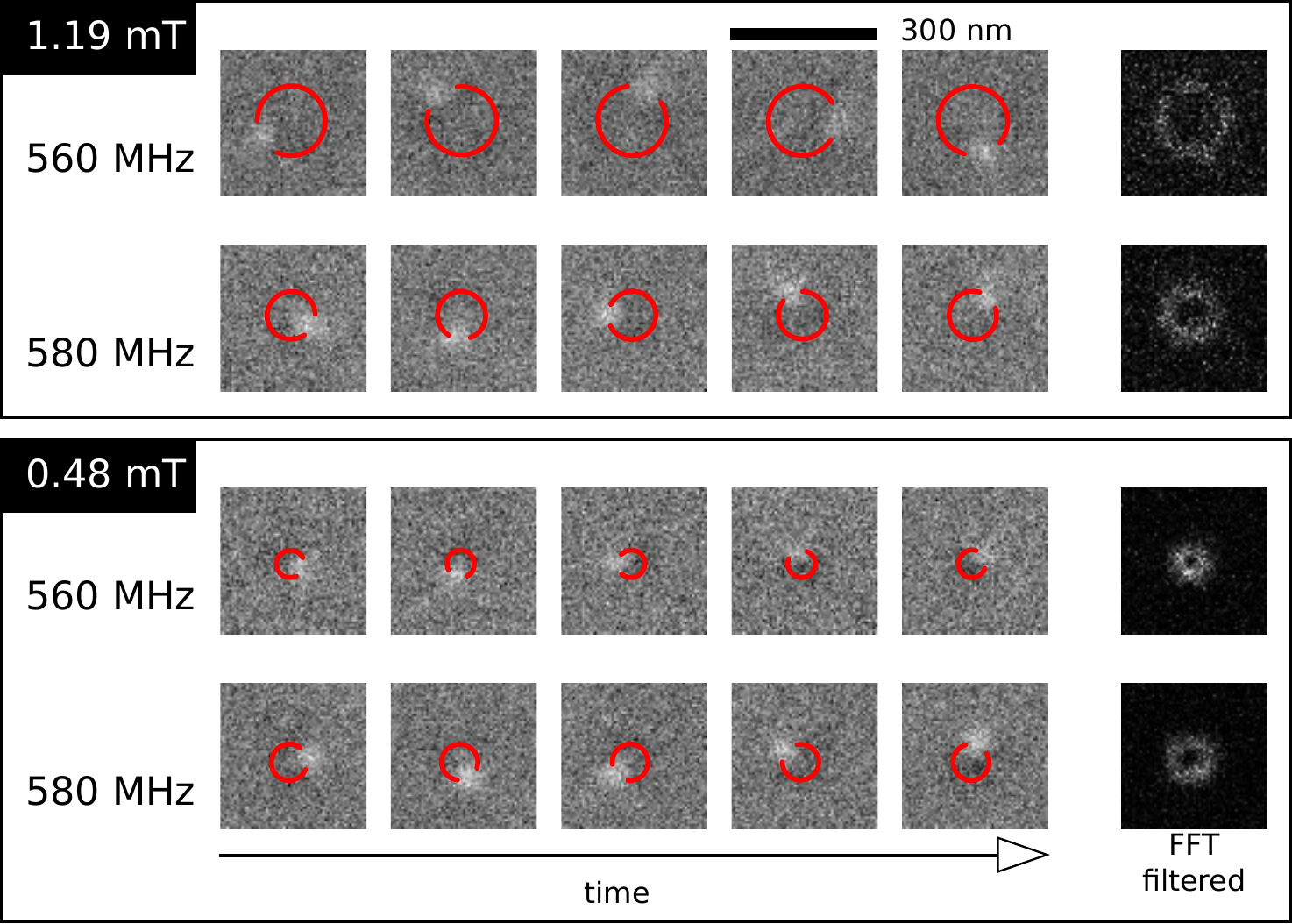}
\caption{Time-resolved STXM images of the vortex core gyration shown at five phases of alternating magnetic fields. The excitation frequencies of 560\,MHz and 580\,MHz are near the gyrotropic resonance frequency. The respective gyration radii can easily be compared by from the rightmost images, which are Fourier-filtered at a harmonic of the excitation frequency. Under the low-field excitation of 0.48\,mT, the gyration radius is larger for the 580\,MHz excitation. On the other hand, at the higher excitation of 1.19\,mT, the resonance has shifted downwards and the gyration radius is now larger at 560\,MHz. The gyrotropic mode is thus anharmonic, as the resonance frequency depends on the excitation strength. \label{fig1}}
\end{figure}

First of all, the experiment shows qualitatively that the resonance frequency decreases when the excitation amplitude is increased. This effect is illustrated in Fig. \ref{fig1}. Five of the 25 recorded frames of the gyrating vortex core are shown for different combinations of excitation frequency and amplitude. The radius of gyration is highlighted in the Fourier-filtered images. These show the magnitude of the image obtained by Fourier-filtering the 25 time-dependent images at a harmonic of the excitation frequency. This technique allows to easily observe the gyration radius with the bare eye, but is not used in the further analysis. It can be observed that at the lowest excitation amplitude the gyration radius is larger at 580\,MHz, while at the highest excitation amplitude the radius is larger at 540\,MHz. The resonance frequency thus shifts down while the field is increased. This frequency ``redshift'' is a nonlinear effect that does appear in a perfect harmonic oscillator.\\

\begin{figure}[!htb]
\centering
\includegraphics[width=0.7\linewidth]{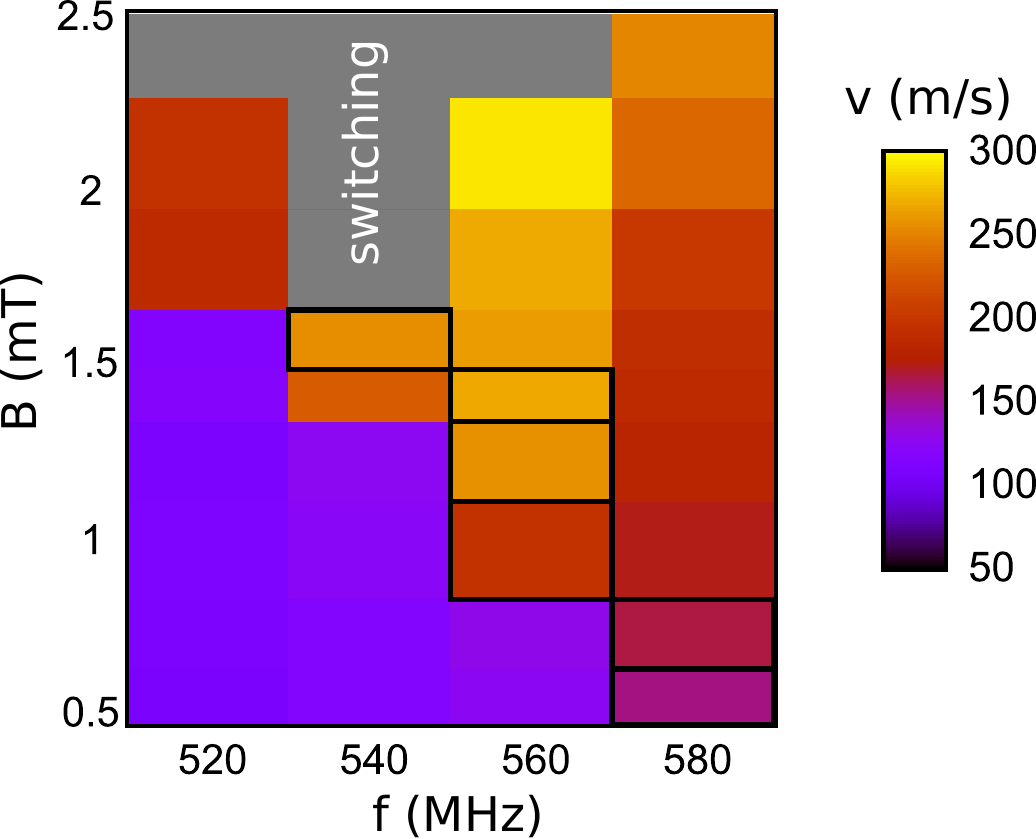}
\caption{Experimentally measured vortex core gyration velocity as a function of excitation frequency and amplitude in a small region around the gyrotropic resonance. The frequencies of maximal gyration are marked with a box, indicating the resonance frequency for each amplitude. This shows that the the gyrotropic resonance frequency shifts down as the amplitude of the excitation field is increased. Above 1.5\,mT, vortex core switching occurs (marked in grey) at the resonance frequency. At even higher fields, switching also occurs at frequencies further away from resonance. \label{fig2}}
\end{figure}

For a more quantitative analysis, the vortex core position in each set of 25 recorded frames was determined by fitting two-dimensional Gaussian profiles to the images. The subtraction of the average image was hereby taken into account. From those positions, the average gyration radius and velocity during one period was deduced. This procedure was repeated at four excitation frequencies near the gyrotropic resonance frequency, as well as at several different excitation amplitudes.\\

A full scan of excitation frequencies and amplitudes is shown in Fig. \ref{fig2}. Here, the average vortex core velocity is shown. It was deduced by multiplying the average gyration radius by the angular velocity of the core, which is fixed by the excitation frequency $f_0$. For each magnetic excitation amplitude, the frequency which lies closest to the resonance, and where the response is thus maximal, has been marked with a box. It can again be observed that the resonance shifts down as the field is increased.\\

At 1.5\,mT and 540\,MHz, the vortex core started toggling back and forth, indicating that the critical switching velocity (about 300\,m/s \cite{guslienko08, yamada07}) was reached. When the excitation amplitude was even further increased, the core also switched at frequencies further away form the resonance. This results in a \emph{switching region} with a typical triangle-like shape, marked in grey. Although it was not yet identified as such, this kind of region is also visible in the FMR spectra in \cite{buchanan07}.\\

\section{Micromagnetic Simulations}\label{simulations}

\subsection{Simulation Method}

The experimental results were compared to micromagnetic simulations of a 500\,nm$\times$\,500\,nm\,$\times$\,50\,nm  Permalloy square structure (saturation magnetization $M_s$=736\x10$^3$\,A/m (taken from \cite{vansteenkiste08}), exchange constant $A$=13\x 10$^{-12}$\,J/m, damping parameter $\alpha$=0.01, anisotropy constant $K_1$=0 and gyromagnetic ratio $\gamma$=2.211\,$\times$\,10$^5$\,m/As). A 2D discretization with 3.9\,$\times$\,3.9\,nm$^2$ finite difference cells was used. The simulations were carried out with our in-house developed \textsc{MuMax} micromagnetic simulator \cite{mumax}. This program runs on Graphical Processing Units (GPUs) rather than on standard CPUs, providing a computational speedup of up to two orders of magnitude. This allowed a large parameter range to be scanned with fine resolution. In total, 8000 separate field/amplitude combinations were simulated. Each simulation started from the relaxed ground state. The field was first applied for 20 periods to allow the system to absorb energy and approach a stationary state. The next 20 periods of the field were used for data-taking. The vortex core position and average velocity were output during those 20 periods. We choose to start each simulation from the ground state and not to, e.g., slowly scan the frequency or field. This avoids possible effects due to a too high scan speed.

\subsection{Simulation Results}

\begin{figure}[!htb]
\centering
\includegraphics[width=0.7\linewidth]{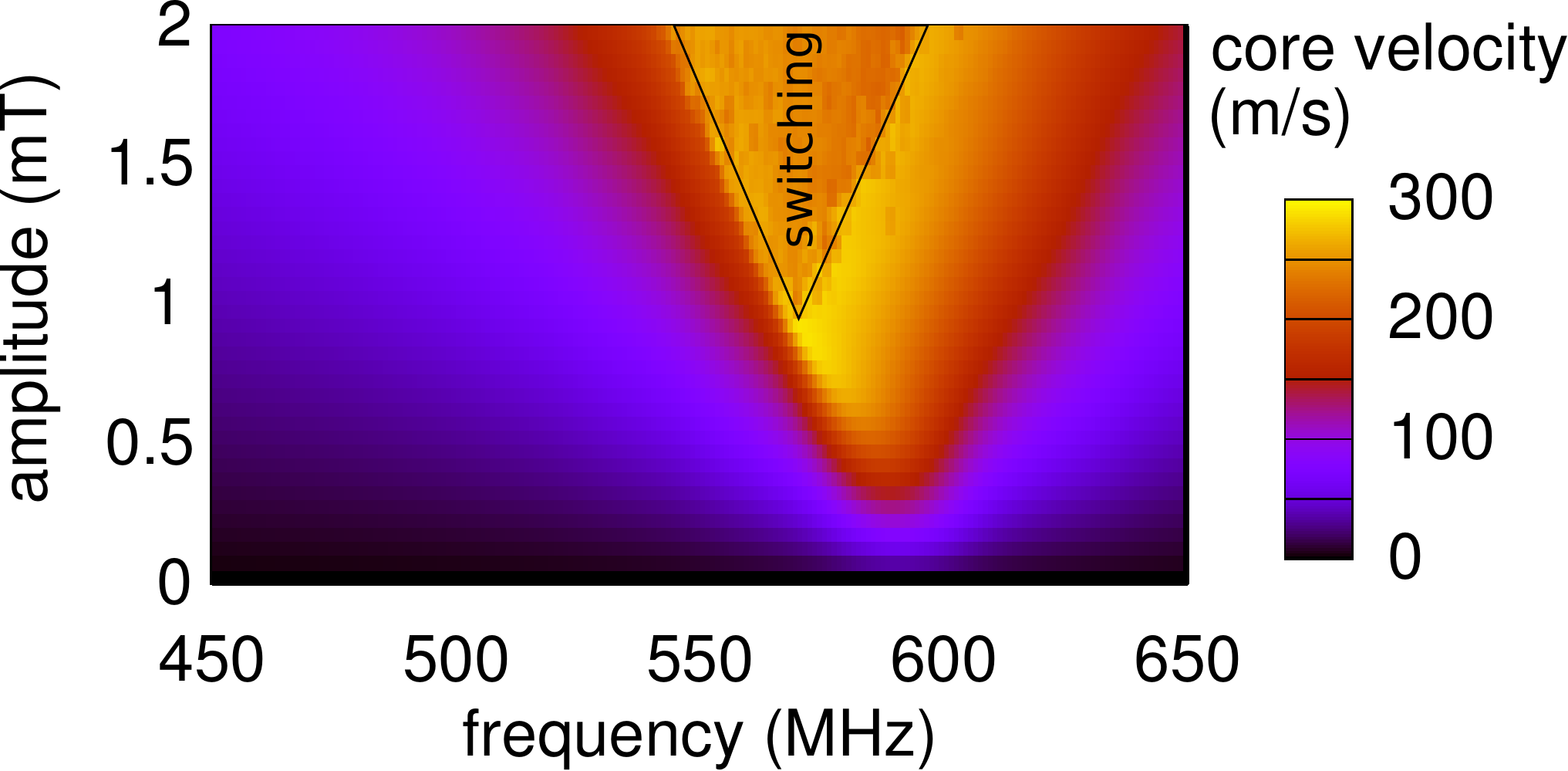}
\caption{Simulated frequency and field dependence of the vortex core gyration velocity for a larger range than the experimental data in Fig. \ref{fig2}. The simulated data qualitatively show the same behavior as the experimental data. The simulations also reveal the triangular region in parameter space where vortex core switching occurs. In this region the average core velocity is lowered due to the continuous switching \cite{yamada07}. Experimentally, the velocity in this region could not be determined.\label{fig3}}
\end{figure}

The simulated gyration velocity as a function of excitation frequency and amplitude is shown in Fig.\,\ref{fig3}. This simulation qualitatively agrees with the experimental results: the frequency redshift at high excitations and the typical shape of the switching region can here also be observed. The small quantitative discrepancies with the experimental results are attributed to uncertainties on the nanostructure size, thickness and material parameters. A more detailed analyses is presented in the next section.

\section{Detailed Analysis}


\subsection{Frequency redshift}

\begin{figure}[!htb]
\centering
\includegraphics[width=1\linewidth]{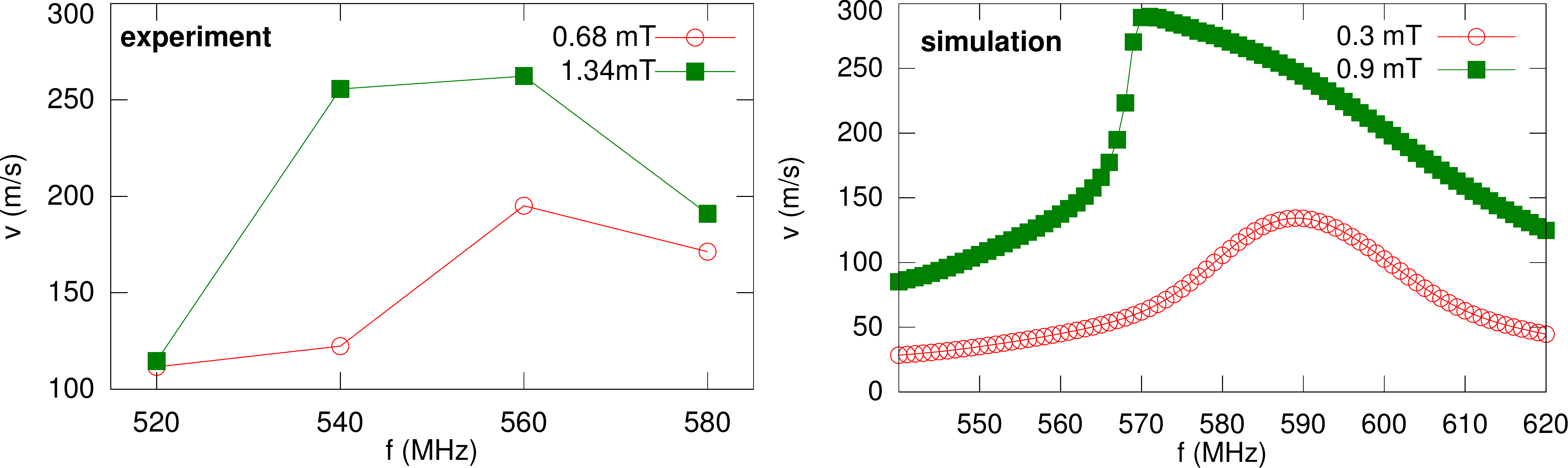}
\caption{Experimental (left) and simulated (right) vortex core velocity $v$ as a function of the excitation frequency $f$ for a low and a high excitation amplitude. The high excitation lies just below the vortex core switching threshold. At high excitation, the peak of the response is clearly redshifted towards a lower frequency. The detailed simulated curve also reveals that the resonance peak shape becomes asymmetric. \label{redshifts}}
\end{figure}

\begin{figure}[!htb]
\centering
\includegraphics[width=0.7\linewidth]{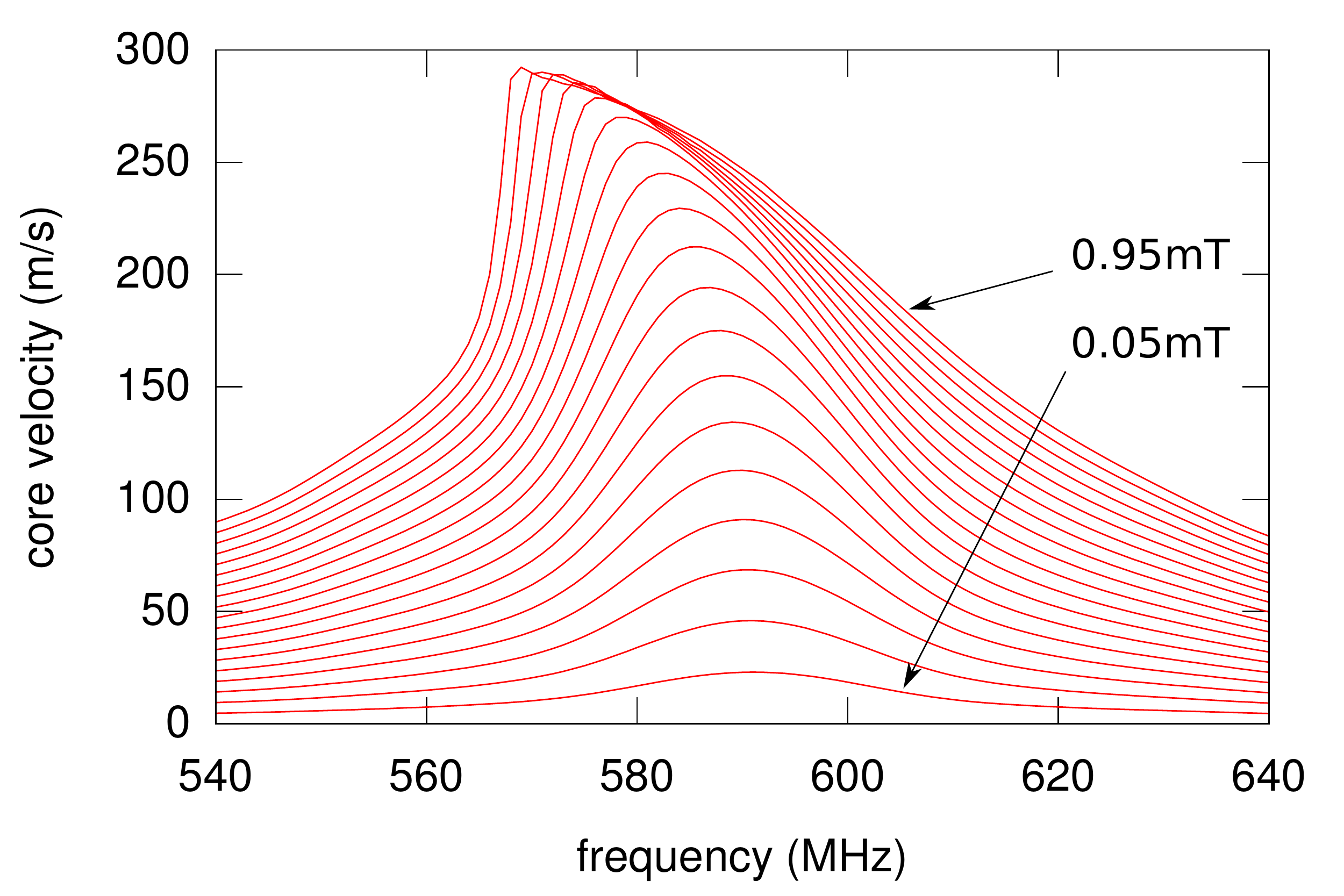}
\caption{Simulated vortex core velocity as a function of the excitation frequency $f$ at various excitation amplitudes until just below the vortex core switching threshold. At high excitations, the resonance peak shifts towards lower frequencies and develops an asymmetric shape. \label{rmsredshift}}
\end{figure}

First of all, we analyse frequency scans at a few fixed excitation amplitudes. Fig. \ref{fig1} already illustrated qualitatively that the resonance frequency drops as the excitation amplitude is increased. This redshift is shown in more detail in Fig. \ref{redshifts}. Here, experimental and simulated frequency scans at two representative excitation amplitudes are presented. One amplitude is chosen well below the vortex core switching threshold, while the second amplitude closely approaches it. It is clear that in both cases high amplitudes cause the resonance peak to become broader and to shift to lower frequencies. The simulated data also reveal that the resonance peak becomes asymmetric --- a typical property of anharmonic oscillators \cite{buchanan07}. Fig. \ref{rmsredshift} shows the resonance peaks for a wider range of excitation amplitudes. Here, the gradual deformation and redshift of the resonance curve at increasingly higher amplitudes can be clearly seen.

\subsection{Amplitude jumps}

\begin{figure}[!htb]
\centering
\includegraphics[width=1\linewidth]{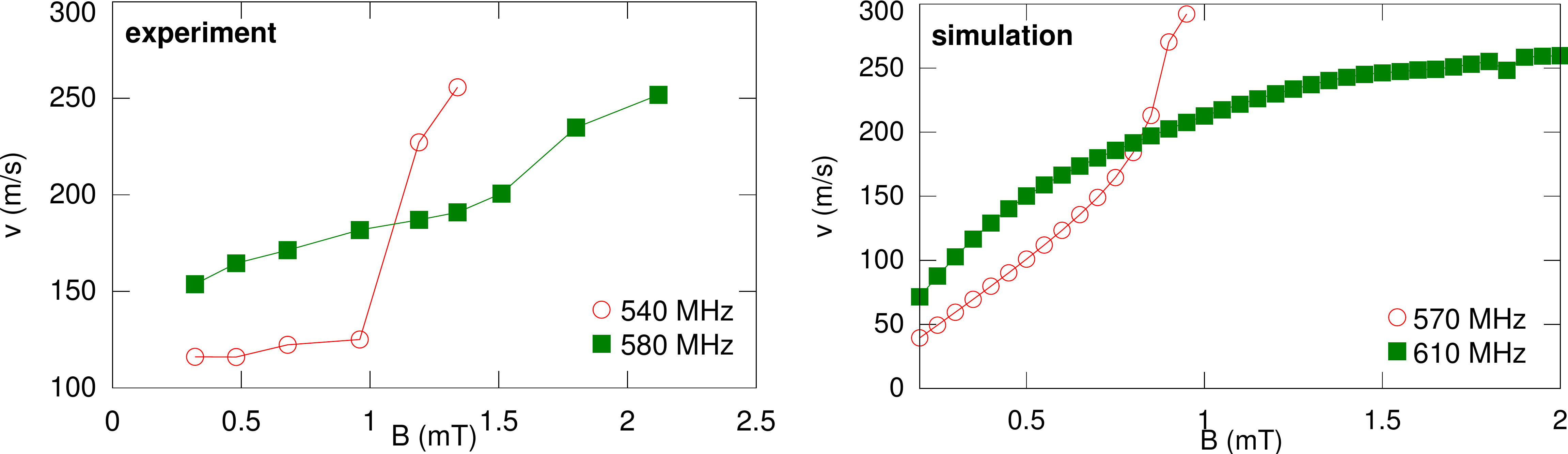}
\caption{Experimental (left) and simulated (right) vortex core velocity $v$ as a function of the excitation amplitude $B$, for a frequency just below and a frequency just above the respective resonances. At the lowest frequency, the velocity increases suddenly around 1mT: while increasing the field amplitude, the resonance redshifts towards the excitation frequency until the resonance frequency suddenly is reached. The higher frequency lies already above the resonance so the resonance peak redshifts away from the excitation frequency as the amplitude increases. Hence, no sharp jump of the response occurs.\label{jumps}}
\end{figure}

Secondly, we analyse amplitude scans at a few fixed frequencies. For a perfect harmonic oscillator, the response (either velocity or displacement) should be proportional to the excitation amplitude. Fig. \ref{jumps} presents experimental and simulated amplitude scans at two fixed frequencies, one slightly below and one slightly above the resonance. At the lowest of these two frequencies, the response exhibits a sharp jump around 1mT, which is clearly non-linear behavior. This behavior can be explained by the redshift as follows: Let us consider a frequency just below the resonance frequency at a relatively small amplitude. By increasing the amplitude, not only the response naturally becomes larger, but also the resonance frequency redshifts towards the excitation frequency, causing an extra increase of the response.  In contrary, this does not happen when a frequency is considered above the (small-amplitude) resonance frequency.  In that case, increasing the amplitude causes the resonance to shift away from the considered excitation and thus counteracts the resonance increase.

\section{Discussion}

Until now, theoretical studies on thinner nanostructures predicted a \emph{blueshift} of the gyrotropic frequency at high excitation amplitudes \cite{lee07-apl, Guslienko2010}. In these studies, relatively thin nanostructures were considered, which allows the vortex core to approach the edges of the magnet. A steepening of the potential well in which the vortex core moves was attributed to the expulsion of the core by the edge, hence the blueshift.  In our case however, the vortex core did not approach the edges of the system. The maximum gyration radius was about 150\,nm (see, e.g., Fig \ref{fig1}) --- much smaller than the lateral size of the magnet. \\

\begin{figure}[!htb]
\centering
\includegraphics[width=0.6\linewidth]{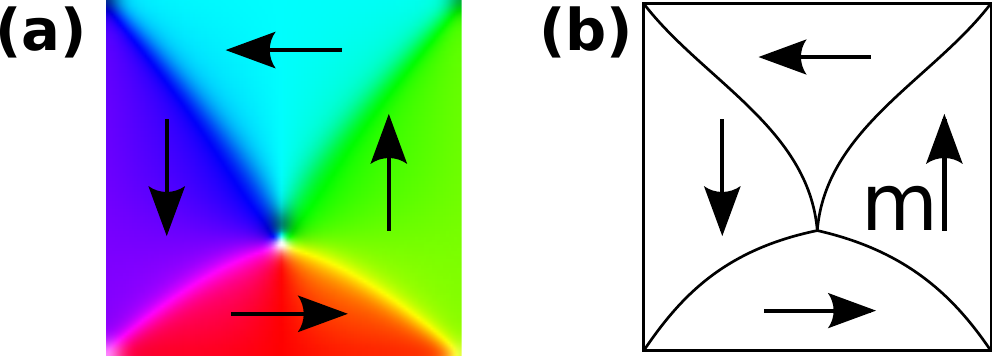}
\caption{\textbf{(a)} Simulated magnetization snapshot at an excitation that leads to gyrotropic redshift. The domain walls have slightly bended, reducing the global dipole moment. As this lowers the dipolar energy, it weakens the potential well in which the vortex core moves ---hence decreasing the eigenfrequency. \textbf{(b)} Exaggerated sketch for clarity. \label{bending}}
\end{figure}

The redshift in this case can be explained by domain wall bending during the vortex movement. Fig. \ref{bending} shows a simulated magnetization snapshot where such bending occurs. It was taken at an excitation just below the threshold for vortex core swichting --- where there is a strong redshift. It can be seen that the domain walls slightly bend so that the total dipole moment is reduced with respect to straight domain walls. Hence the magnetostatic energy of the potential well and thus the resonance frequency is reduced \cite{kruger07}.\\

Nevertheless, this effect is not specific to square-shaped nanostructures. Micromagnetic simulations on disks (not presented here) also revealed a redshift. Although domain walls are absent in that case, the magnetization adapted in a similar way, reducing the global dipole moment.\\

\section{Conclusion}

We have performed a detailed analysis of the non-linearity of the magnetic vortex gyration in a 50\,nm thick, 500\,nm wide Permalloy nanostructure. The study has shown that a frequency \emph{redshift} occurs at high excitation amplitudes. By using time-resolved magnetic STXM microscopy the gyrotropic motion could be imaged directly, allowing for a straightforward interpretation of the results. STXM also allows to detect the vortex core polarization and switching. In this way, the region in frequency/amplitude space where the vortex core switches could be determined. This was not possible in earlier FMR studies like \cite{buchanan07}.  Extensive micromagnetic simulations were carried out on Graphical Processing Units and support the experimental results.\\

\section*{Acknowledgments}
Financial support by The Institute for the promotion of Innovation by Science and Technology in Flanders (IWT-Flanders) and by the Research Foundation Flanders (FWO-Flanders) through the research grant 60170.06 are gratefully acknowledged. The Advanced Light Source is supported by the Director, Office of Science, Office of Basic Energy Sciences, of the U.S. Department of Energy. We would also like to cordially thank André Drews for the fruitful discussions, including the explanation of the redshift by domain wall bending.

\section*{References}
\bibliography{bibliography}

\end{document}